\documentclass{llncs}

\usepackage{amsmath}
\usepackage{amsfonts}
\usepackage{amssymb}
\usepackage{graphicx}
\usepackage{epstopdf}
\usepackage[noend]{algorithm2e}

\newcounter{lemmas}
\newcounter{definitions}

\newtheorem{FRM1_lemma}[lemmas]{Lemma}
\newtheorem{FRM2_lemma}[lemmas]{Lemma}
\newtheorem{jsliding_def}[definitions]{Definition}
\newtheorem{jdisjoint_def}[definitions]{Definition}

\begin{document}
\title{Generic Subsequence Matching Framework:\\Modularity, Flexibility, Efficiency}

\author{
David Novak \and Petr Volny \and Pavel Zezula}
\institute{ Masaryk University, Brno, Czech Republic\\
\email{[david.novak\,|\,xvolny1\,|\,zezula]@fi.muni.cz}}

\maketitle
\begin{abstract}
Subsequence matching has appeared to be an ideal approach for solving many
problems related to the fields of data mining and similarity retrieval. It has
been shown that almost any data class (audio, image, biometrics, signals) is or
can be represented by some kind of time series or string of symbols, which can
be seen as an input for various subsequence matching approaches. The variety of
data types, specific tasks and their partial or full solutions is so wide that
the choice, implementation and parametrization of a suitable solution for a
given task might be complicated and time-consuming; a possibly fruitful
combination of fragments from different research areas may not be obvious nor
easy to realize. The leading authors of this field also mention the
implementation bias that makes difficult a proper comparison of competing
approaches. Therefore we present a new generic Subsequence Matching Framework
(SMF) that tries to overcome the aforementioned problems by a uniform frame that
simplifies and speeds up the design, development and evaluation of subsequence
matching related systems. We identify several relatively separate subtasks
solved differently over the literature and SMF enables to combine them in
straightforward manner achieving new quality and efficiency. This framework can
be used in many application domains and its components can be reused
effectively. Its strictly modular architecture and openness enables also
involvement of efficient solutions from different fields, for instance efficient
metric-based indexes.

This is an extended version of a paper published on DEXA 2012.
\end{abstract}




\section{Introduction}
Majority of the data being produced in current digital era is in the form of
\emph{time series} or can be transformed into sequences of numbers. This concept
is very natural and ubiquitous: audio signals, various biometric data, image
features, economic data, etc. are often viewed as time series and need to be also
organized and searched in this way. 

One of the key research issues drawing a lot of attention during the last two
decades is the \emph{subsequence matching problem}, which can be basically
formulated
as follows: Given a query sequence, find the best-matching subsequence from the
sequences in the database. Depending on the specific data and application, this
general problem has many variants -- query sequences of fixed or variable size,
data-specific definition of sequence matching, requirement of dynamic time
warping, etc. Therefore, the effort in this research area resulted in many
approaches and techniques -- both, very general and those focusing on a
specific fragment of this complex problem. 

The leading authors in this field identified two main problems that limit the
comparability and cooperation potential of various approaches: the \emph{data
bias} (algorithms are often evaluated on heterogeneous datasets) and the
\emph{implementation bias} (the implementation of the specific technique can strongly
influence experiment results)~\cite{Keogh2002}. The effort to overcome the
data bias is expressed by founding a common set of data collections
\cite{KeoghUCR} which is publicly available and that should be used by any
consequent research in this area. However, the implementation bias lingers,
which also obstructs a straightforward combination of compatible approaches whose
interconnection could be very efficient.


Analysis of this situation brought us to conclusion that there is a need for a
unified environment for developing, prototyping, testing, and combination of
subsequence matching approaches. In this paper we propose such generic
subsequence matching framework (SMF), namely:
\begin{itemize}
	\item we overview and decompose the state-of-the-art approaches and
		techniques for subsequence matching and we identify several common
		sub-problems that these approaches deal with in various ways
		(Section~\ref{sec:overview});
	\item we propose general architecture of SMF that should fulfill our
		targets and we describe our implementation of the framework
		(Section~\ref{sec:framework}); power of SMF is demonstrated by
		elegant realization of several variants of fundamental subsequence matching algorithms;
	\item we describe real applications with different requirements for
		subsequence matching that can be simply implemented with the aid
		of our framework (Section~\ref{sec:demos}).
\end{itemize}
The paper concludes in Section~\ref{sec:conclusions} by future directions
that cover possible performance boost enabled by a straightforward cooperation
of our framework with advanced distance-based indexing and
searching technologies~\cite{Zezula2006,NBZ2010mindex}.

\section{Time Series Processing Overview}
\label{sec:overview}
The field opening paper by Faloutsos et al. \cite{Faloutsos1994} introduced a
subsequence matching application model that has been used ever since 
only with smaller modifications. The model can be summarized in the following
four steps that should be adopted by a subsequence matching application:
\begin{description}
\item[slicing] of the time series sequences (both data and query) into shorter
	subsequences (of a fixed length),
\item[transforming] each subsequence into lower dimension, 
\item[indexing] the subsequences in a multi-dimensional index structure,
\item[searching] in the index with a distance measure that obeys the 
	lower bounding lemma on the transformed data.
\end{description}
In the original work~\cite{Faloutsos1994}, this approach was demonstrated on
a subsequence matching algorithm that used the \emph{sliding window} approach to
slice the indexed data and \emph{disjoint window} for the query.
The Discrete Fourier Transformation (DFT) was used for dimensionality reduction
and the data was indexed using the minimum bounding rectangles in
R-tree~\cite{Guttman1984}.  The Euclidean distance was used for searching since it 
satisfies the lower bounding lemma on data transformed by DFT.

\begin{figure*}[tbp]
\centering
\includegraphics[width=\textwidth]{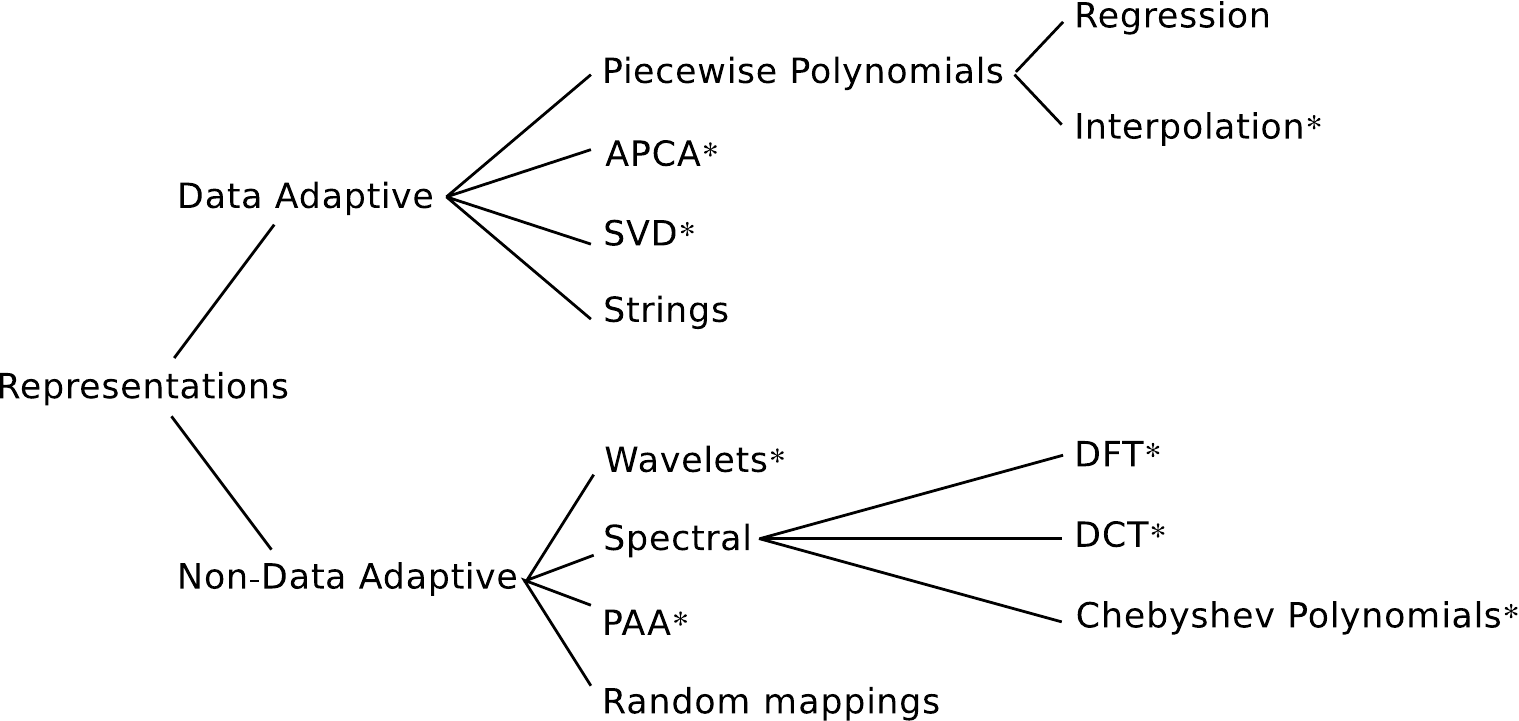}
\caption{Basic hierarchy of representations. Those with asterisk after the name allows lower bounding.}
\label{fig:representations}
\end{figure*}

Many works followed this model contributing to it by using other time series
representations, more sophisticated distance functions and corresponding lower
bounding mechanisms. In the rest of this section, we provide a brief overview of
the key approaches in this field.

\subsection{Data Representation}
\label{sec:representation}
The way in which the data is represented may be crucial for both efficiency and
effectiveness of the whole system. 
Naturally, one can work directly with the original data, but this \emph{raw
data} is typically very large and it would require great computing
infrastructure to process such data fast enough. Therefore, a lot of effort
has been aimed at
finding the best possible approximation of the time series that would reduce the
dimensionality of the data on one hand and preserve the main features of
the data on the other. 

Figure~\ref{fig:representations} depicts the
most significant approaches used for reducing time series data representation: 
\textit{Discrete Fourier Transformation} (DFT)~\cite{Faloutsos1994},
\textit{Discrete Cosine Transform} (DCT)~\cite{Korn1997},  
\textit{Chebyshev Polynomials} (CHEB)~\cite{Cai2004},
\textit{Discrete Wavelet Transform} (DWT)~\cite{Chan1999},
\textit{Piecewise Aggregate Approximation} (PAA)~\cite{Keogh2000},
\textit{Single Value Decomposition} (SVD)~\cite{Korn1997},
\textit{Adaptive Piecewise Constant Approximation} (APCA)~\cite{Chakrabarti2002}.
On the higher level, we distinguish two groups of approaches: \emph{data
adaptive} and \emph{non-data adaptive}. An important feature for any
representation is the ability to be indexed because proper indexing can
dramatically improve performance of the whole retrieval process. Members of the
latter group sometimes struggle with indexing~\cite{Yi2000}. Another
desirable property of the data representation is the possibility to
perform \textit{lower bounding}, which can again improve the performance of the
subsequence matching process. 

\subsection{Distance computation}
\label{sec:distances}
\begin{figure}[tbp]
\centering
\includegraphics[width=0.75\textwidth]{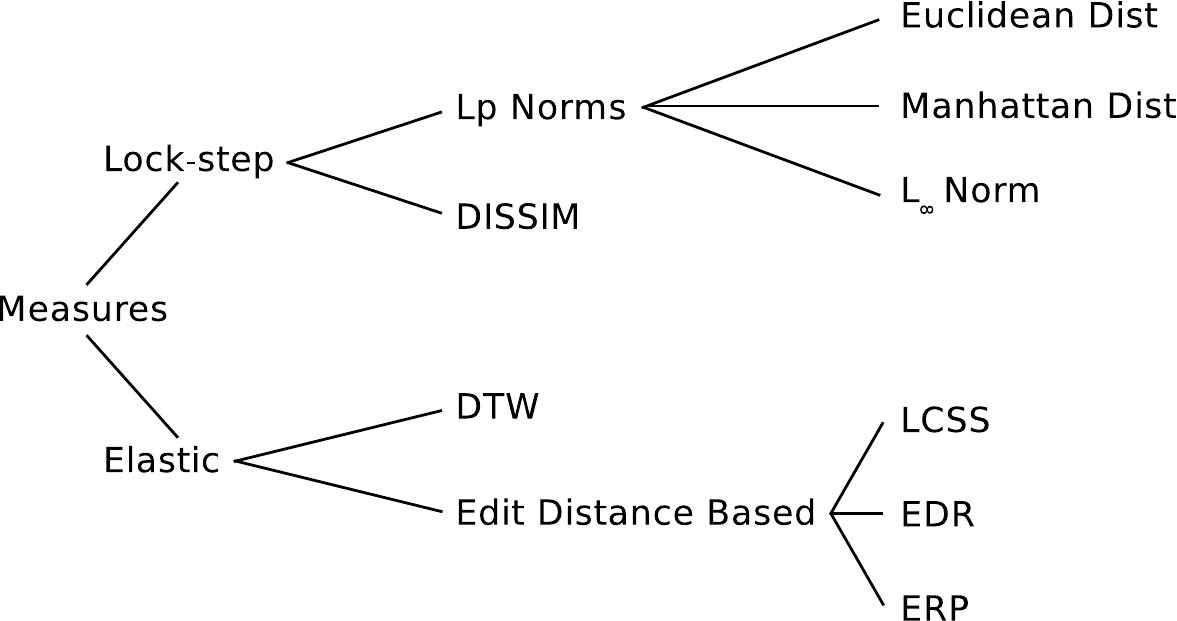}
\caption{Basic hierarchy of distance functions.}
\label{fig:measures}
\end{figure}

The notion of similarity between two sequences is typically expressed by
a \emph{distance measure} (distance function). The most frequently used functions are depicted in
Figure~\ref{fig:measures}. The lock-step measures, like Euclidean distance, are
usually relatively cheap to compute but they 
lack the robustness against even the basic data transformations. On the other hand, more sophisticated
dynamic programming methods like Dynamic Time Warping (DTW) allow shifts on the time axis and serve
better to applications like speech recognition or query by humming. It is also
important to pair the 
data representation and the distance function wisely in order to satisfy the lower bounding
lemma~\cite{Faloutsos1994}. Distance functions can also differ by their input
domain. Some functions are defined on
continuous domains (real numbers) and some work on strings (sequences of symbols from
a finite alphabet). The most common distance measures are:
\textit{Euclidean Distance} (ED)~\cite{Faloutsos1994},
\textit{Dynamic Time Warping} (DTW)~\cite{Sakoe1978,Kim2001,Keogh2004},
\textit{Edit Distance with Real Penalty} (ERP)~\cite{Chen2004},
\textit{Edit Distance on Real Sequence} (EDR)~\cite{Chen2005},
\textit{Longest Common Subsequence} (LCSS)~\cite{Vlachos2002}.

\subsection{Subsequence Matching Approaches}
\label{sec:approaches}
In order to build a subsequence matching application, number of questions has to be answered: What
kind of data will be used and what is its dimensionality? What is the volume of
the data? Do we need the warping ability? Are
approximate answers acceptable? etc. On the basis of the
answers, one can make proper design decisions like what representation and distance measure to use,
which storage index to employ, etc.

The above mentioned work by Faloutsos et al.~\cite{Faloutsos1994} encouraged
many new approaches that followed his work. Moon et al.~\cite{Moon2001}
suggested dual approach for slicing and indexing sequences to the one used in
\cite{Faloutsos1994}. This \emph{DualMatch} uses the
sliding windows for queries and disjoint windows for data sequences to reduce
the number of windows that are indexed.  DualMatch was followed by the
generalization of windows creation method 
called GeneralMatch \cite{Moon2002}.
Another significant leap forward was made by the effort of Keogh et al. in their work about
exact indexing of Dynamic Time Warping \cite{Keogh2004}. They introduced a similarity measure that
is relatively easy to compute and it lower-bounds the expensive DTW function. This approach was further
enhanced by improving I/O part of the subsequence matching process using Deferred Group Subsequence
Retrieval introduced in \cite{Han2007}.
Moreover, many new distance functions and data representations were
introduced as we outlined in the previous sections.

If we focus on the performance side of the system, we have to employ
enhancements like indexing, lower bounding, window size optimization, reducing
I/O operations or approximate queries. Lots of approaches for building
subsequence matching applications often use the very same techniques for
solving common sub-tasks included in the whole retrieval process and changes
only some parts with some novel approach. This leads to implementing the same
parts of the process like DFT or DWT repeatedly which leads to the phenomenon of
the \textit{implementation bias}~\cite{Keogh2002}. The modern subsequence
matching approaches \cite{Han2007,Keogh2004} employ many smaller tasks in the retrieval
process that solve sub-problems like optimizing I/O operations. Implementations of routines that
solve such sub-problems should be reusable and employable in similar approaches.
This led us to think about the whole subsequence matching process as a chain of
subtasks, each solving a small part of the problem. We have observed that many
of the published approaches fit into this model and their novelty is often only
in reordering, changing or adding new subtask implementation into the chain.

\section{Subsequence Matching Framework}
\label{sec:framework}
In this section, we describe the general Subsequence Matching Framework (SMF) 
that is currently available under GPL license at
\url{http://mufin.fi.muni.cz/smf/}. The framework can be perceived on the
following two levels that should, naturally, coincide: 
\begin{itemize}
	\item on the \emph{conceptual level}, the framework is composed of mutually
		cooperating modules, each of which solves a specific sub-task, and
		these modules are cooperating within specific subsequence
		matching algorithms; 
	\item on the \emph{implementation level}, the framework defines the
		functionality of individual module types and their communication
		interfaces; a subsequence matching algorithm is then
		implemented as a \emph{skeleton} that combines modules in a specific way
		and this skeleton can be filled with actual module implementations.
\end{itemize}
In Section~\ref{sec:modules}, we describe the common sub-problems (sub-tasks)
that we identified in the field and we define corresponding types of modules (conceptual level).
Further, in Section~\ref{sec:algorithms}, we justify our approach by describing
fundamental subsequence algorithms in terms of our modules and we present a
straightforward implementation of these algorithms within SMF.
Section~\ref{sec:implementation} is devoted to details about implementation of
the framework including an example of a configuration file by which one can
create a brand new algorithm only by exchanging specific modules in a
text configuration file.


The key term in the whole framework is, naturally, a \emph{sequence}. As we want
to keep the framework as general as possible, we do not lay practically any
restrictions on the components of the sequence -- it can be integers, real
numbers, vectors of numbers, or any more sophisticated structures. The
sequence similarity functions are defined relatively independently of specific
sequence type (see Section~\ref{sec:implementation}).
In the following, we will use the notation summarized in Table~\ref{tab:notation}. 

\newcommand{\len}[1]{#1.\mathit{len}}
\newcommand{\id}[1]{#1.\mathit{id}}
\newcommand{\parent}[1]{#1.\mathit{pid}}
\newcommand{\offset}[1]{#1.\mathit{offset}}

\begin{table}[tbp]
\begin{center}
	\caption{Notation used throughout this paper.}
	\label{tab:notation}
\begin{tabular}{|l||l|}
\hline
\bfseries Symbol\,&\,\bfseries Definition \\
\hline
\hline
$S[k]$ & the $k$-th value of the sequence $S$ \\ \hline
$S[i:j]$ & subsequence of $S$ from $S[i]$ to $S[j]$, inclusive \\ \hline
$\len{S}$ & the length of sequence $S$ \\ \hline
$\id{S}$ & the unique identifier of sequence $S$ \\ \hline
$\parent{S'}$ & if $S'$ is subsequence of $S$ then $\parent{S'}=\id{S}$ \\ \hline
$\offset{S'}$ & if $S'=S[i:j]$  then $\offset{S'}=i$ and $\len{S'}=j-i+1$  \\ \hline
$D(Q,S)$ & distance between two sequences $Q$ and $S$ \\ \hline
\end{tabular}
\vspace*{-15pt}
\end{center}
\end{table}

\subsection{Common Sub-problems: Modules in SMF}
\label{sec:modules}
Studying the field of subsequence matching, we identified several common
sub-problems addressed by a number of approaches in some sense. Specifically, we
can see the following sub-tasks that correspond to isolated modules in our framework.


\subsubsection{Data Representation (Data Transformer Module)}
\begin{figure*}[tbp]
\centering
\includegraphics[width=0.95\textwidth]{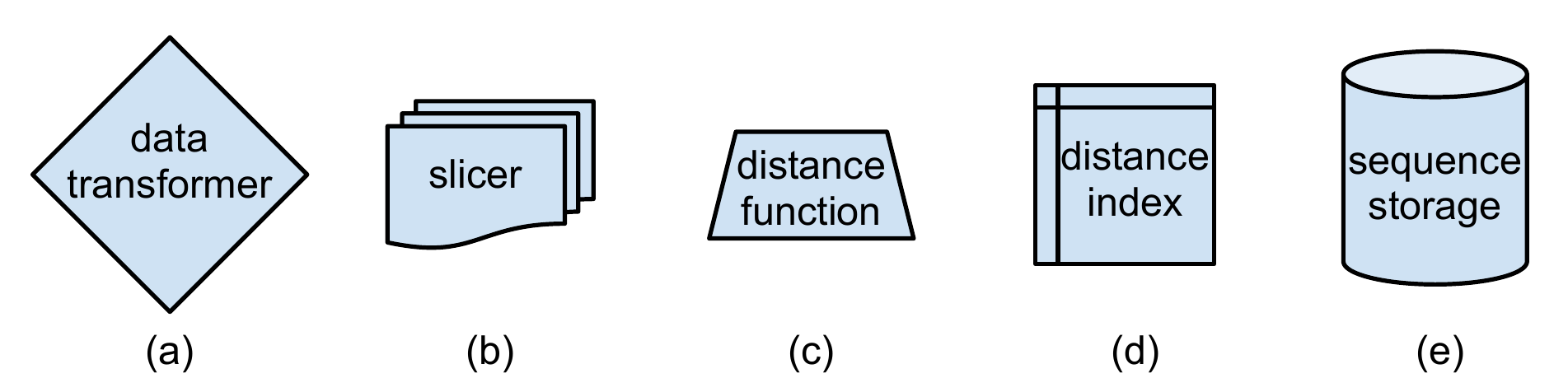}
\caption{Types of SMF modules and their notation.}
\label{fig:modules}
\end{figure*}

The raw data sequences entering an application are often transformed into other
representation which can be motivated either by simple dimensionality
reduction (DFT, DWT, SVD, PAA)~\cite{Faloutsos1994,Chan1999,Korn1997,Keogh2000}
or also by extracting some important characteristics that should improve the
effectiveness of the retrieval~\cite{Perng2000} (see
Section~\ref{sec:representation} for details). In either case, the general task can
be defined simply as follows: \emph{Transform given sequence $S$ into
another sequence $S'$.} We will use the symbol in Figure~\ref{fig:modules} (a)
for this \emph{data transformer} module. The following table summarizes information about this
module and gives a few examples of specific approaches implementing this
functionality.
\begin{center}
	\begin{tabular}{c|l}
	\bfseries data transformer\, &\,\bfseries transform sequence $S$ into sequence
	$S'$ \\ \hline
	DFT & apply the DFT on a sequence of real numbers $S$~\cite{Faloutsos1994} \\
	PAA & apply the PAA on a sequence of real numbers $S$~\cite{Keogh2000} \\
	Landmarks & extract \emph{landmarks} from a sequence $S$~\cite{Perng2000}
\end{tabular}
\end{center}

\subsubsection{Windows and Subwindows (Slicer Module)}
Majority of the subsequence matching approaches partitions the data and/or query
sequences into subsequences of, typically, fixed
length (windows)~\cite{Faloutsos1994,Moon2001,Moon2002,Han2007}. Again, this
task can be isolated, well defined, and the implementation can be reused in many
variants of subsequence matching algorithms. Partitioning a sequence $S$, each
resulting subsequence $S'=S[i:j]$ has $\parent{S'}=\id{S}$, $\offset{S'}=i$, and
$\len{S'}=j-i+1$. The module will be denoted as in
Figure~\ref{fig:modules} (b) and its description and specific
examples are as follows:
\begin{center}
	\begin{tabular}{c|l}
	\bfseries sequence slicer\, &\,\bfseries partition $S$ into list
	of subsequences $S'_1,\ldots,S'_n$ \\ \hline
	disjoint slicer & partition $S$ disjointly into subsequences of length $w$~\cite{Faloutsos1994} \\
	sliding slicer & use sliding window of size $w$ to partition $S$~\cite{Faloutsos1994} \\
\end{tabular}
\end{center}


\subsubsection{Distance Functions (Distance Module)}
As analyzed in Section~\ref{sec:distances}, there emerged a high number of
specific distance functions $D$ that can be evaluated between two sequences $S$
and $T$. The intention of SMF is to partially separate the distance functions from the
data and to use the specific distance function as a parameter of the algorithm
(see Section~\ref{sec:implementation} for details on implementation of the this independence).
Of course, it is the matter of configuration to use appropriate function for
respective data type, e.g. to preserve the lower bounding property.
The distance functions symbol is in Figure~\ref{fig:modules} (c) and it can be
summarized as follows:
\begin{center}
	\begin{tabular}{c|l}
	\bfseries distance function\, &\,\bfseries evaluate dissimilarity of
	sequences $S$, $T$ \\ \hline
	$L_p$ metrics & evaluate distance $L_p$ on equally long number sequences \\ 
	DTW & use DTW on any pair of number sequences $S$, $T$~\cite{Sakoe1978} \\
	ERP & calculate Edit distance with Real Penalty on $S$, $T$~\cite{Chen2004} \\
	LB\_PAA, LB\_Keogh\, & measures that lower bound the DTW~\cite{Keogh2004}
\end{tabular}
\end{center}

\subsubsection{Efficient Indexing (Distance Index Module)}
An efficient subsequence-matching algorithm typically employs an index to
efficiently evaluate distance-based queries on the stored (sub-)sequences using
the query-by-example paradigm (QBE). Again, we see the choice of the specific index as a relatively
separate component of the whole algorithm and thus as an exchangeable module. Also,
we see a space for improvement in boosting the efficiency of this component in
future. We denote this module as in Figure~\ref{fig:modules} (d):
\begin{center}
	\begin{tabular}{c|l}
	\bfseries distance index\, &\,\bfseries evaluate efficiently distance-based QBE queries
	\\ \hline
	R-Tree family & index sequences as $n$-dimensional spatial data \\ 
	$i$SAX tree & use a symbolic representation of the
	sequences~\cite{Shieh2008,Camerra2010} \\
	metric indexes & index and search the data according to mutual
	distances~\cite{Zezula2006} \\
\end{tabular}
\end{center}

\subsubsection{Efficient Aligning (Sequence Storage Module)}
The approaches that use sequence slicing typically also need to store the
original whole sequences. The slice index (for window size $w$) returns a set
of candidate subsequences $S'$, $\len{S'}=w$ each matching some query
subsequence $Q'$ such that $\len{Q'}=w$. If the query sequence $Q$ is actually
longer than $w$, the subsequent task is to align $Q$ to corresponding
subsequence
$S[i:(i+\len{Q}-1)]$ where $i=\offset{S'}-\offset{Q'}$ and 
$\id{S}=\parent{S'}$. To do this aligning for each $S'$ in the candidate
set may be very demanding. For smaller datasets, this can be done in
memory with no special treatment, but more advanced approaches are profitable on
disk~\cite{Han2007}. We will call this module \emph{sequence storage}
(Figure~\ref{fig:modules} (e)) and it is specified as follows:
\begin{center}
	\begin{tabular}{c|l}
	\bfseries sequence storage\, &\,\bfseries store sequences $S$ and
	return $S[i:j]$ for given $\id{S}$ \\ \hline
	hash map & basic hash map evaluating queries one by one \\
	deferred retrieval & deferred group sequence retrieval (I/O
	efficient)~\cite{Han2007}\\
\end{tabular}
\end{center}

\subsection{Subsequence Matching Strategies in SMF}
\label{sec:algorithms}
\begin{figure*}[tbp]
\centering
\includegraphics[width=\textwidth]{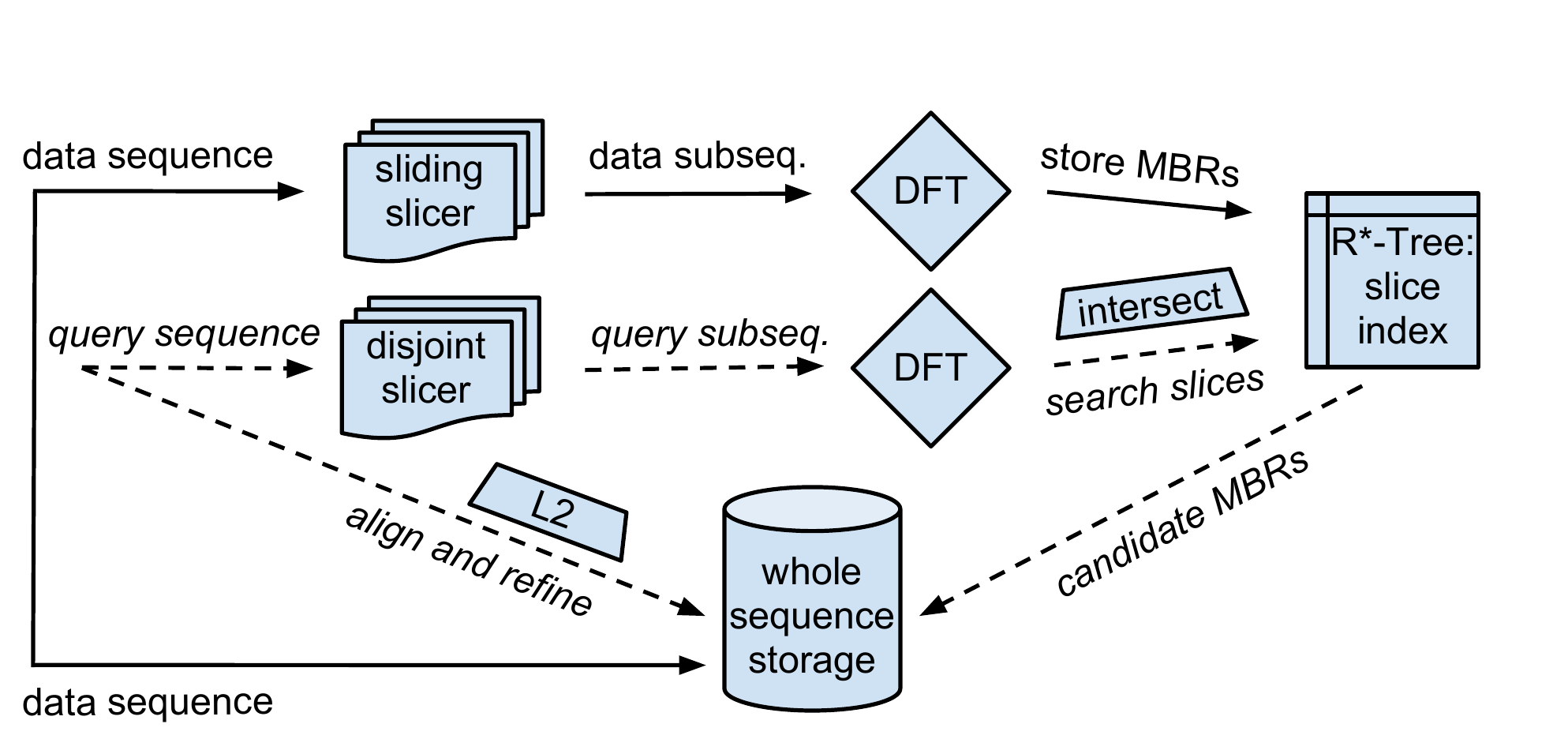}
\caption{Schema of the fundamental subsequence matching algorithm~\cite{Faloutsos1994}.}
\label{fig:faloutsos}
\end{figure*}

Staying at the conceptual level, let us have a look at the whole subsequence
matching algorithms and their composition from individual modules introduced
above. As an example, we take again the fundamental 
algorithm~\cite{Faloutsos1994} for general subsequence matching of queries
$Q$, $\len{Q}\geq w$ for an established window size $w$. The schema of a
slight modification of this algorithm is in Figure~\ref{fig:faloutsos}.  The
solid lines correspond to data insertion 
and the dash lines (with italic labels) correspond to the query processing. 

A data sequence $S$ is first partitioned by the sliding window
approach (\emph{sliding slicer} module) into slices $S'=S[i:(i+w-1)]$, these are
transformed by Discrete Fourier Transformation
(\emph{data transformer} module DFT), and the Minimum Bounding Rectangles
(MBR) of these transformed slices are stored in an R$^*$-tree storage
(\emph{distance index} module); the
original sequences $S$ is also stored (\emph{whole sequence storage} module).
Processing a subsequence query, the query sequence $Q$ is partitioned using the
\emph{disjoint slicer} module, each slice $Q'=Q[i:(i+w-1)]$ is transformed by DFT and
it is searched within the slice index (using $L_2$ distance or a simple
binary function \emph{intersect}). For each of the returned candidate subsequences
$S'$, a query-corresponding alignment $S[i:(i+\len{Q}-1)]$ is retrieved
from the \emph{whole storage} (see above for details) and the candidate set is
refined using $L_2$ \emph{distance} $D(Q,S[i:(i+\len{Q}-1)])$.

%

Preserving the skeleton of an algorithm (module types and their cooperation),
one can substitute individual modules with other compatible modules obtaining
a different processing efficiency or even a fundamentally different algorithm.
For instance, swapping the sliding and disjoint slicer
modules practically results in the DualMatch approach~\cite{Moon2001}.
Exchanging the R$^*$-tree index for a metric-based index like
the M-Index~\cite{NBZ2010mindex} could possibly improve the efficiency for
larger dimensionality of the stored slices (especially
if the M-Index could use approximate evaluation strategy). In general, a fast and
straightforward alternation of modules is very helpful when seeking the best
solution for a particular task and data collection and tuning the performance of
this solution.


\subsection{Implementation of SMF}
\label{sec:implementation}
The SMF was not implemented from scratch but with the aid of 
framework MESSIF~\cite{Batko2007}. The MESSIF is a collection of Java
packages supporting mainly development of metric-based search approaches. From
MESSIF, the SMF uses especially the following functionality:
\begin{itemize}
	\item encapsulation of the concept of data objects and distances,
	\item implementation of the queries and query evaluation process,
	\item distance based indexes (building, querying),
	\item configuration and management of the algorithm via text \emph{config files}.
\end{itemize}

\subsubsection{Data Independence} 
The \emph{sequence} is in SMF handled very generally; it is defined as an
interface which requires that each specific sequence type (e.g. a simple float
sequence) must, among other, specify the distance between two sequence components
$d(S[i], S'[j])$. For number sequences, this distances could be, naturally,
absolute value of differences $d(S[i],S'[j])=|S[i]-S'[j]|$, but one can imagine complex sequence
components, for instance vectors where $d$ could be an $L_p$ metric. 
Implementation of a sequence distance $D(S,S')$ (for instance, DTW) then treats
$S$ and $S'$ only as general sequences with component distance $d$ and, thus,
this implementation can be independent of specific sequence type.

\subsubsection{Module Implementation}
The SMF module types specified in Section~\ref{sec:modules} are typically
implemented as Java interfaces and the specific modules as Java classes. 
The interface specifies prototypes of the methods (inputs and outputs) that the
specific module must provide.

\subsubsection{Algorithm Implementation}
Realizing a specific algorithm, one must implement its skeleton -- module types,
their connections, and all algorithm-specific operations. This skeleton is
compiled within the SMF package together with all the interfaces and module
implementations. 
Then, the algorithm is instantiated only by means of a text configuration file -- module
types required by the algorithm skeleton are filled by specific modules. 

\bigskip

\begin{figure*}[tbp]
\centering
\includegraphics[width=0.8\textwidth]{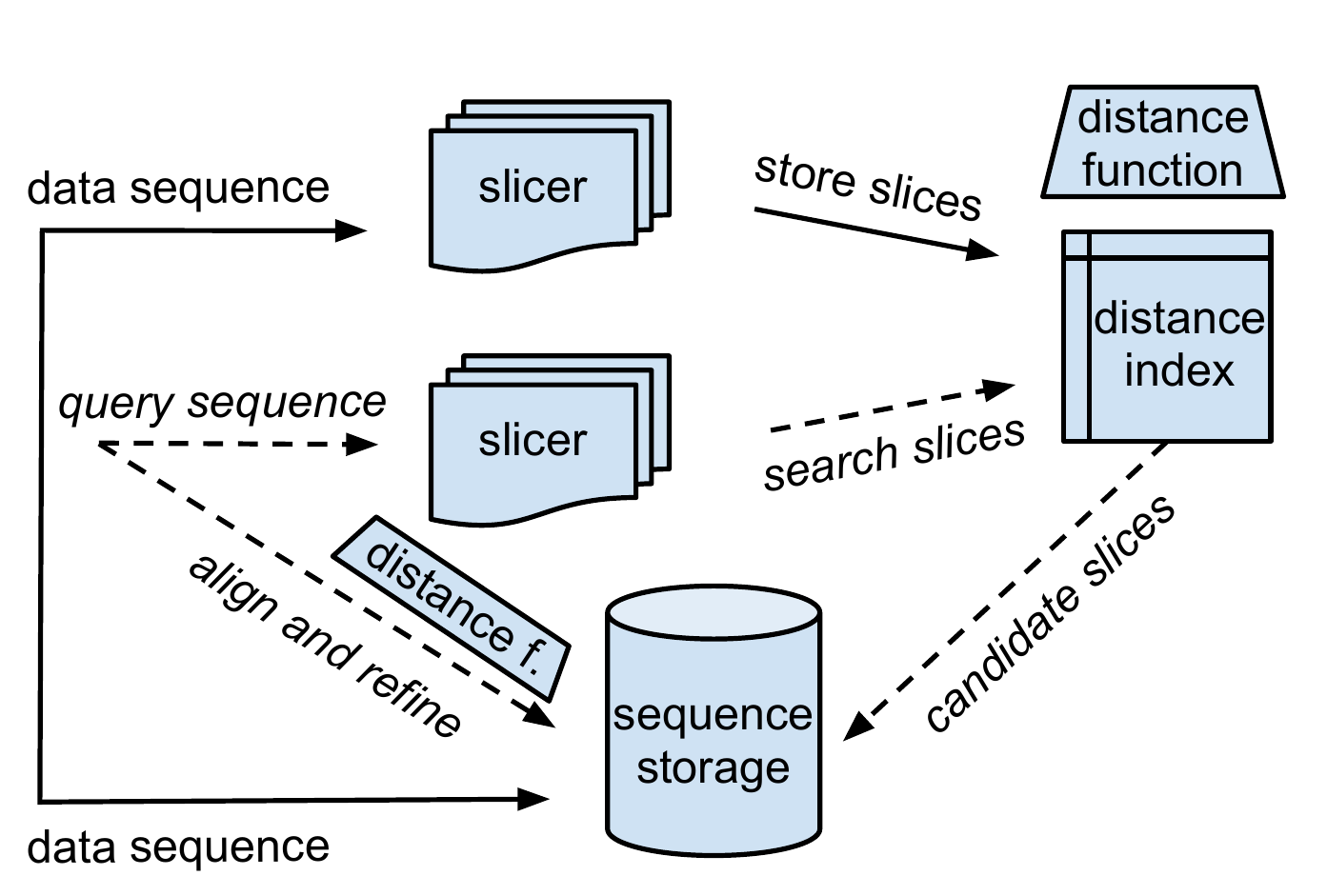}
\caption{Skeleton of {\tt VariableQueryAlgorithm}, a simple subsequence matching
algorithm with variable length query.}
\label{fig:simple}
\end{figure*}

Let us describe this principle on an example of a simple algorithm for general
subsequence matching with variable query length -- see
Figure~\ref{fig:simple} for the schema of the skeleton. This schema
uses two \emph{slicer} modules, one \emph{distance index} with a \emph{distance
function}, and a \emph{sequence storage} for the whole sequences (again, with a
\emph{distance function}). 

In order to run a specific algorithm, we have to instantiate these module
types by specific modules. Figure~\ref{fig:config} shows a part of a
SMF configuration file that starts such algorithm. The syntax of these files is
taken from the MESSIF framework and it is quite intuitive. On
the first two lines, the \emph{sliding slicer} module is created by an action
called \verb#namedInstanceAdd# which creates an instance \verb#slidingSlicer# 
of class \verb#smf.modules.slicer.SlidingSlicer# (with parameter $w$); the
\emph{disjoint slicer} is created accordingly. Then, the
instance of \emph{distance index} is created; it is a self-standing
\emph{algoritm}, namely a metric index M-Index~\cite{NBZ2010mindex} (we assume
that the instance \verb#mIndex# has been already created). In this
example, the \emph{sequence storage} is instantiated as a simple memory storage
(\verb#seqStorage#).

\begin{figure}[tbp]
\begin{verbatim}
slidingSlicer = namedInstanceAdd
slidingSlicer.param.1 = smf.modules.slicer.SlidingSlicer(<w>)

disjointSlicer = namedInstanceAdd
disjointSlicer.param.1 = smf.modules.slicer.DisjointSlicer(<w>)

index = namedInstanceAdd
index.param.1 = smf.modules.index
                           .ApproximateAlgorithmDistanceIndex(mIndex)

seqStorage = namedInstanceAdd
seqStorage.param.1 = smf.modules.seqstorage.MemorySequenceStorage()

startSearchAlg = algorithmStart
startSearchAlg.param.1 = smf.algorithms.VariableQueryAlgorithm
startSearchAlg.param.2 = smf.sequence.impl.SequenceFloatL2
startSearchAlg.param.3 = seqStorage
startSearchAlg.param.4 = index
startSearchAlg.param.5 = slidingSlicer
startSearchAlg.param.6 = disjointSlicer
startSearchAlg.param.7 = <w>
\end{verbatim}
\caption{\label{fig:config} Example of a SMF configuration file}
\end{figure}

Finally, the actual \verb#VariableQueryAlgorithm# is started passing the
created module instances as parameters to the skeleton. The \verb#param.2# of
this action specifies that this particular algorithm instance requires sequences
of floating point numbers and will compare them by Euclidean distance. Such SMF
configuration files are supported directly by the MESSIF
framework that enables an efficient management of the running algorithm.

\section{Use Cases}
\label{sec:demos}
One of the motivations for building the SMF framework were several applications that
all need subsequence matching but are relatively heterogeneous. Let us
demonstrate three such applications that differ in data type, distance
function, query specification, and requirements for indexing efficiency; all
of them can be straightforwardly implemented using SMF.

\subsection{General Subsequence Matching System}
\begin{figure*}[tbp]
\centering
\includegraphics[width=\textwidth]{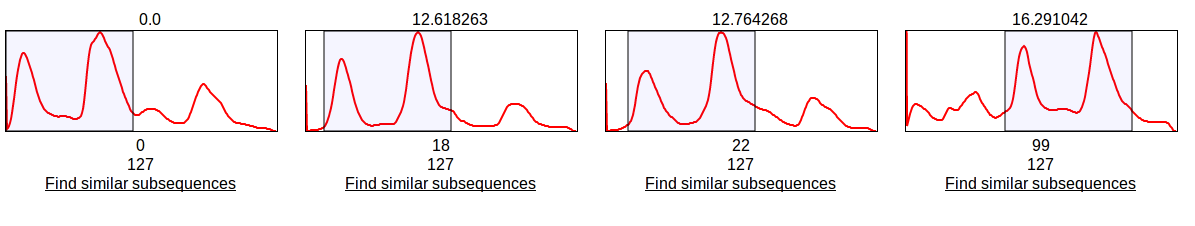}
\caption{General subsequence matching system with variable query length.}
\label{fig:screenshot}
\end{figure*}

The first demonstration is a general subsequence matching system for queries
with variable length. The demo is built on a small collection of simple
time-series data and it uses the algorithm described in
Section~\ref{sec:implementation} (Figures~\ref{fig:simple}
and~\ref{fig:config}). Figure~\ref{fig:screenshot} shows screenshot of the GUI
of this demo: The user can specify a subsequence (offset and width) of
the query sequence and after clicking ``Find similar subsequences'', the
most similar ones (according to Euclidean distance) are located (their distances
are above the answer series). The demo
is publicly available at \url{http://mufin.fi.muni.cz/subseq/}. The
characteristics of this simple demo can be summarized as follows:
\begin{center}
\begin{tabular}{c|l}
\bfseries data type & series of real numbers \\ \hline
\bfseries similarity measure & Euclidean distance \\ \hline
\bfseries query type & subsequence query with variable length (longer than $w$) \\ \hline
\bfseries requirements & no additional requirements
\end{tabular}
\end{center}


\subsection{Gait Recognition}
\begin{figure*}[htb]
\centering
\includegraphics[width=0.53\textwidth]{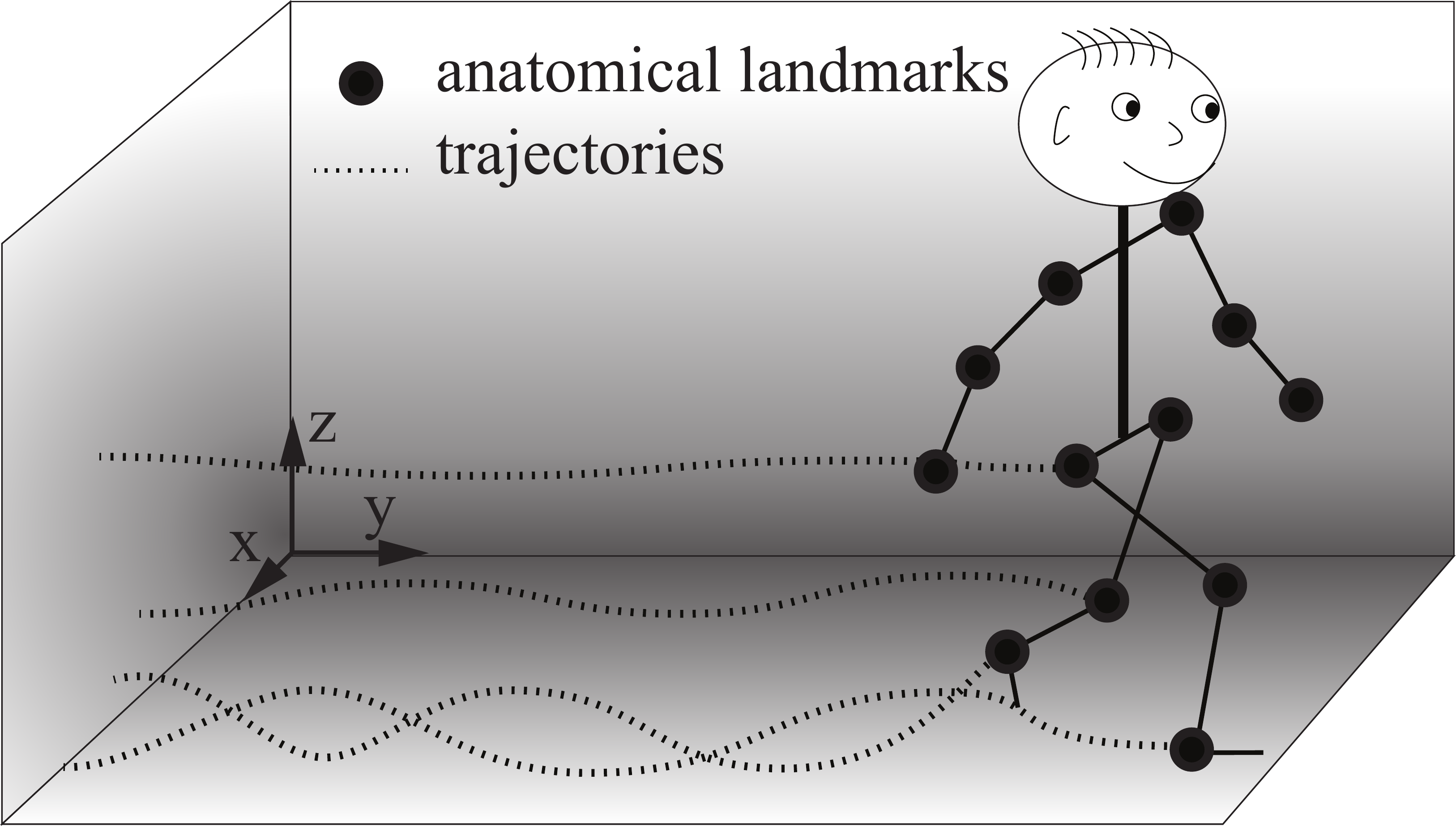}
\includegraphics[width=0.43\textwidth]{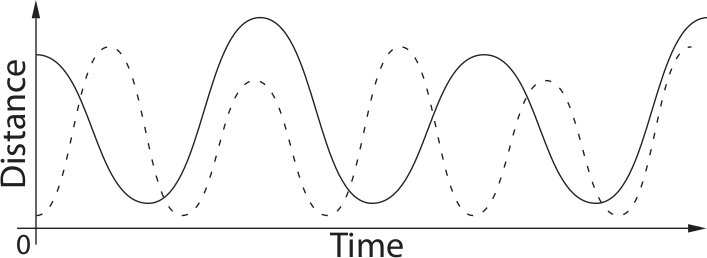}
\caption{Gait recognition based on 3D coordinates of anatomical landmarks.}
\label{fig:gait}
\end{figure*}

The biometrics form a wide application area that often manages data in the
form of sequences. Among others, recognition of people according to their
\emph{gait} characteristics is currently a on the increase. There are several
fundamental approaches to this problem and one of them works directly with 3D coordinates of
anatomical landmarks of human body. The feasibility of this approach is growing
as the hardware that can extract such 3D positions is more and more
mature and available~\cite{Bhanu10Gait}. The left image in
Figure~\ref{fig:gait} sketches the principle. The trajectories are further
processed and the development of mutual distances between various anatomical
landmarks is studied~\cite{Valcik12} -- see time-series in the right part of Figure~\ref{fig:gait}.
Currently, the SMF is used for research in this area:
\begin{center}
\begin{tabular}{c|l}
\bfseries data type & series of real numbers or of number vectors \\ \hline
\bfseries similarity measure & various $L_p$ metrics, DTW, special measures \\ \hline
\bfseries query type & subsequence query with fixed or variable length \\ \hline
\bfseries requirements & data and similarity flexibility, high performance in future
\end{tabular}
\end{center}

\subsection{Unsupervised Spoken Term Detection}
The Automatic Speech Recognition (ASR) is an extremely important area for
human-computer interaction and the fundamental problem in this field is Spoken Term
Detection (STD). Besides well-developed complex approaches based on
specific language models, the pattern recognition based on \emph{unsupervised
methods}, that focus on situations when no linguistic corpus is available,
represent quite recent stream of research~\cite{Volny2011}. One of
the studied approaches uses \emph{posteriorgram templates}~\cite{Hazen2009}
extracted using a phonetic recognizer. The DTW measure is often used with this
data~\cite{Hazen2009}, but we would like to evaluate other options like ERD and
ERP comparing the results. Again, the SMF is an ideal platform for this
research and future applications:
\begin{center}
\begin{tabular}{c|l}
\bfseries data type & series of probabilities or of probability vectors \\ \hline
\bfseries similarity measure & DTW, ERD, ERP; special distances between components \\ \hline
\bfseries query type & subsequence query with variable length \\ \hline
\bfseries requirements & data and similarity flexibility, high performance in future
\end{tabular}
\end{center}

\section{Conclusions and Future Work}
\label{sec:conclusions}
The data in the form of sequences are all around us in various forms and extensive
volumes. The research in the area of subsequence matching has been very
intensive resulting in many partial or full solutions in various sub-areas of
the field. In this work, we identified several sub-tasks that circulate over the
field and are tackled within various subsequence matching approaches and
algorithms. 

We present a generic subsequence matching framework (SMF) that brings the option
of choosing freely among the existing partial solutions and combining them in
order to achieve ideal solutions for heterogeneous requirements of different
applications. Also, this framework overcomes the often mentioned implementation
bias present in the field and it enables a straightforward utilization of
techniques from different areas, for instance advanced metric indexes. We
describe SMF on conceptual and implementation levels and present several
examples and demonstration applications from diverse fields. The SMF is
available under GPL license at \url{http://mufin.fi.muni.cz/smf/}.

The architecture of the SMF framework is strictly modular and thus one of
natural directions of future development is implementation of other modules.
Also, we will develop SMF according to requirements emerging from continuous 
research streams that utilize SMF. Finally and most importantly, we would like
to contribute to the efficiency of the subsequence matching systems by
involvement of advanced metric indexes. We believe in a positive impact of
such cooperation of these two research fields that were so far evolving
relatively separately.


\section*{Acknowledgments}
This work was supported by national research projects GACR 103/10/0886, and GACR
P202/10/P220.

\bibliographystyle{splncs}
\bibliography{library}

\end{document}